\documentclass[iop,jcp,preprint,amsmath,amssymb,superscriptaddress]{revtex4-1}

\usepackage[utf8]{inputenc}
\usepackage[T1]{fontenc}

\usepackage[version=3]{mhchem}
\usepackage{hyperref}
\usepackage{graphicx}
\graphicspath{{figures/}}

\usepackage{dcolumn}% Align table columns on decimal point
\usepackage{bm}% bold math
\usepackage{color}
\begin{document}

\title{The relevance of structural variability in the time-domain for computational reflection anisotropy spectroscopy at solid--liquid interfaces}

\author{Justus Leist}

\affiliation{Universität Ulm, Institute of Theoretical Chemistry, Ulm, Germany}

\author{Jongmin Kim}

\affiliation{Universität Tübingen, Institute of Physical and Theoretical Chemistry, Tübingen, Germany}
\affiliation{Universität Ulm, Institute of Theoretical Chemistry, Ulm, Germany}

\author{Holger Euchner}

\affiliation{Universität Tübingen, Institute of Physical and Theoretical Chemistry, Tübingen, Germany}

\author{Matthias M. May}
\email{matthias.may@uni-tuebingen.de}
%\altaffiliation{}
\affiliation{Universität Tübingen, Institute of Physical and Theoretical Chemistry, Tübingen, Germany}
\affiliation{Universität Ulm, Institute of Theoretical Chemistry, Ulm, Germany}

\date{\today}

\begin{abstract}

In electrochemistry, reactions and charge-transfer are to a large extent determined by the atomistic structure of the solid-liquid interface. Yet due to the presence of the liquid electrolyte, many surface-science methods cannot be applied here. Hence, the exact microscopic structure that is present under operating conditions often remains unknown. Reflection anisotropy spectroscopy (RAS) is one of the few techniques that allow for an \textit{in operando} investigation of the structure of solid-liquid interfaces. However, an interpretation of RAS data on the atomistic scale can only be obtained by comparison to computational spectroscopy. While the number of computational RAS studies related to electrochemical systems is currently still limited, those studies so far have not taken into account the dynamic nature of the solid-liquid interface. In this work, we investigate the temporal evolution of the spectroscopic response of the Au(110) missing row reconstruction in contact with water by combining \textit{ab initio} molecular dynamics with computational spectroscopy. Our results show significant changes in the time evolution of the RA spectra, in particular providing an explanation for the typically observed differences in intensity when comparing theory and experiment. Moreover, these findings point to the importance of structural surface/interface variability while at the same time emphasising the potential of RAS for probing these dynamic interfaces.
\end{abstract}

\maketitle

\section{Introduction}
Solid-liquid interfaces are crucial for many technologically relevant processes, ranging from energy storage in batteries to hydrogen production via water splitting \cite{Chu_roadmap_solar_water_splitting_2017}. Charges need to be transferred over this interface to drive an electrochemical reaction, yet the structure and the potential distribution can vary significantly both in the time and space domain, but also as a function of an applied external potential \cite{May_coelec_photoelectrosynthetic_interfaces_2022}. Hence, a detailed understanding of these interfaces is of utmost importance for systematic improvements of electrochemical devices. Significant experimental and computational efforts are therefore dedicated to unravelling the structure of electrochemical interfaces \cite{Magnussen_atomic-scale_understanding_elchem_interfaces_2019,Zhang_electrochemical_systems_finite_field_MD_2020,May_coelec_photoelectrosynthetic_interfaces_2022}. However, most experimental techniques applicable to solid--liquid interfaces under \textit{operando} conditions are either restricted in structural or temporal resolution, which means that the availability of information on the atomistic scale under realistic electrochemical conditions is limited. Here, electrochemical reflection anisotropy spectroscopy (RAS) is an emerging optical method in the field, allowing for a non--destructive investigation of crystalline surfaces and interfaces providing insights in their atomistic structure \cite{Guidat_EC-RAS_review_2023}. 

RAS measures the difference in reflectivity for two orthogonal polarisation directions, thus yielding information on the anisotropy of the investigated surface. This anisotropy can originate (almost) exclusively from the investigated crystal surface, as is for instance the case for the (110) surface of fcc metals or the (100) surface of III-V semiconductors in the zincblende structure. The combination of structural and temporal resolution of RAS has made it a popular tool to follow and control epitaxial growth processes \cite{Haberland_RAS_setup_1999}. Yet due to its near-normal incidence reflection geometry, RAS can also be straightforwardly applied in electrochemical environments to probe the anisotropy and hence the atomistic structure at the electrochemical solid-liquid interface under applied potentials. This means that electrodes with anisotropic crystal surfaces can be investigated under operating conditions, allowing to relate particular features in a cyclic voltammogram to an \textit{in operando} measured spectroscopic response. The observed spectra can then ideally be connected to particular structural modifications such as reconstructions or the adsorption of certain atoms and molecules on the surface \cite{Loew_InP_RAS_2022,Guidat_EC-RAS_review_2023}. Early electrochemical RAS studies used gold as electrode in aqueous electrolytes \cite{Mazine_electrochemical_RAS_Au110_1999, Smith_electrochemical_oxidation_Au110_RAS_2007,Smith_Au110_electrochemical_RAS_2016}. As Au is not prone to corrosion in mild electrolyte concentrations and moderate potential windows, structural changes can be assumed to be driven by the applied potential in a reversible manner \cite{Kolb_surface_reconstructions_Au_elchem_1986,Tidswell_XRD_Au100_in_alkaline_and_acidic_electrolytes_1993}. 

While RA spectroscopy allows to monitor changes on surfaces and interfaces, the interpretation of such changes in terms of the underlying atomic structure is non--trivial. In fact, such an interpretation can only be obtained by comparison between experimental and computational spectra or the tight correlation with complementary experimental methods. While the latter is often challenging or not possible at all for electrochemical systems, the former requires a rather evolved computational effort, depending on the level of theory. Before density functional theory (DFT) calculations of large supercells became affordable, early studies used perturbative approaches \cite{Wang_calc_RAS_Au_110_1990}. The currently most well-established approach for such a calculation is an excited state calculation on a random phase approximation (RPA) or even Bethe-Salpeter equation (BSE) level on top of a DFT ground state \cite{Schmidt_large_scale_simulation_of_surface_optical_spectra_2006}. The existing computational studies on RAS so far focused on fixed structures after geometry optimisation \cite{Hogan_optical_fingerprints_Au_Si_surface_2013,May_water_adsorption_calc_RAS_2018,Loew_InP_RAS_2022}, thus lacking any information on the structural interface dynamics inherent at an electrochemical interface. These dynamics in RAS are expected to qualitatively differ from infrared spectroscopy probing molecular vibrations, where methods are already established to derive experimental signatures from molecular dynamics (MD) simulations
\cite{Roberts_structural_rearrangements_H2O_2D-IR_spectroscopy_2009,Zhang_first-principle_analysis_IR_water_2010}: For most systems it is safe to assume that the bulk electrolyte is isotropic and hence the anisotropy confined to the interface and furthermore, experimental realisations of RAS are typically operating at different energies, in the visible regime ($>1.5$\,eV). 

In this paper, we therefore address the question on how structural dynamics at the solid--liquid interface impact reflection anisotropy spectroscopy. As established model system, we investigate the missing row reconstruction of the Au(110) surface in contact with water. We find significant fluctuations of spectral features over the time of the molecular dynamics trajectory which should be taken into account to realistically model electrochemical interfaces for comparison with experiment.

\section{Methods}
\subsection{Reflection Anisotropy Spectroscopy}

In experimental RA spectroscopy, linearly polarised light at near-normal incidence is used to investigate the surface/interface of interest. The difference in reflectivity along two orthogonal crystal directions in the surface plane is measured and analyzed as a function of the energy of the incoming photons. The observed reflectivity difference is normalised by the overall reflectivity, thus resulting in the following expression:

\begin{equation}
\frac{\Delta r}{r} = 2\frac{r_x - r_y}{r_x + r_y}; r \in \mathbb{C}
\label{eq:expRAS}
\end{equation}

Here, it has to be noted that the typically determined quantity in experiment, $\Delta r/r$, refers to the reflectivity, which can be related to the reflectance via $\mbox{Re}(\Delta r/r)\approx 1/2(\Delta R/R)$ \cite{Schmidt_computational_RAS_review_2005}.
In the case of optically isotropic bulk materials, the signal originates exclusively from the (near) surface region. Especially then, RAS is a highly surface-/interface-sensitive probe that allows to identify subtle structural changes, for instance during an electrochemical process.

From a theoretical perspective, the optical properties of a given system are determined by its dielectric function, such that accessing RA spectra by computational approaches necessitates the determination of the surface dielectric function, as described in the following equation~\cite{Selci_1987, Hogan_2018}:

\begin{equation}
{{\frac{\Delta R}{R}}}= \frac{4\pi d}{\lambda}[A\Delta\varepsilon_{s}''-B\Delta\varepsilon_{s}'],
\label{eqcomras}
\end{equation}
with $\Delta\varepsilon_{s}''$ and $\Delta\varepsilon_{s}'$ corresponding to real and imaginary part of the surface dielectric anisotropy, respectively, $d$ to the layer thickness, and $\lambda$ to the photon wavelength.

Finally, $A$ and $B$ are related to real and imaginary part of the complex bulk dielectric function, $\varepsilon_{b} = \varepsilon_{b}'+ i\varepsilon_{b}''$:

\begin{equation}
A= \Re\Bigl\lbrack\frac{1}{\varepsilon_{b} - 1}\Bigr\rbrack = \frac{\varepsilon_{b}'-1}{(1-\varepsilon_{b}')^{2}+(\varepsilon_{b}'')^2},
\label{eqcomras_A}
\end{equation}
and
\begin{equation}
B=-\Im\Bigl\lbrack\frac{1}{\varepsilon_{b} - 1}\Bigr\rbrack = \frac{\varepsilon_{b}''}{(1-\varepsilon_{b}')^{2}+(\varepsilon_{b}'')^2}. 
\label{eqcomras_B}
\end{equation}

These quantities need to be computed starting from a ground-state electronic structure which is typically provided on a DFT-level of theory.

\subsection{Computational details}

The Au(110) surface and its interaction with water was modelled by periodic density functional theory using the CP2K code package\cite{Kuehne_cp2k_2020}. For all calculations, the Gaussian-and-plane-waves approach was employed, making use of the Goedecker-Teter-Hutter (GTH) pseudopotentials for describing the electron core interaction \cite{Goedecker_separable_Gaussian_pseudopotentials_1996}. A cutoff of 400\,Ry and a relative cutoff of 60\,Ry were selected. Exchange and correlation were described by the generalised gradient approximation in the formulation of Perdew, Burke and Ernzerhof (PBE) \cite{Perdew_GGA_made_simple_1996}. Additionally, van-der-Waals interactions were accounted for by the Grimme D3 correction \cite{Grimme_DFT-D_dispersion_correction_2010}. 

\begin{figure}[t!]
    \centering
    \includegraphics[width=0.5\textwidth]{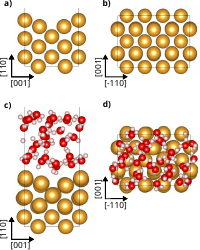}

    \caption{Side and top view of the Au(110) missing row reconstruction in vacuum are shown in panels a) and b), while the setup including water is shown in panels c) and d).}
    \label{fig:Au110}
\end{figure}

Before constructing the Au surface, the unit cell of bulk Au was optimised to determine the lattice parameter. Then, a $2 \times4 $ Au(110) surface with 6 layers was constructed and a row of Au atoms in the top layer was removed to create the \textit{missing row}-reconstruction (see Fig.~\ref{fig:Au110}) \cite{Wang_calc_RAS_Au_110_1990}. After adding 15\,{\AA} of vacuum on top of the surface, the slab geometry was optimised with the bottom 2 layers fixed. The calculations were performed under periodic boundary conditions, using a $5 \times 5 \times 1$ k-point mesh constructed by the Monkhorst-Pack scheme. To allow for the investigation of the solid-liquid interface, 5 water layers -- corresponding to 30 water molecules -- were added on top of the optimised surface to fill the vacuum, resulting in a density close to 1\,g/cm$^3$. 

To obtain the corresponding minimum energy configuration, the geometry of the Au-water system was optimised as well. This configuration was then used as starting structure for subsequent \textit{ab initio} molecular dynamics (AIMD) simulations. For this purpose a canonical ensemble with a temperature of 300 K, controlled by a Nosé–Hoover thermostat, was set up. The AIMD simulations were run for 5000 steps with a timestep of 0.5 fs to equilibrate the system. From the following 5000 steps, 20 configurations (\textit{i.e.} every 250 steps) were extracted and selected for excited state calculation. To determine the corresponding RA spectra, the RAS module as implemented in the Yambo code\cite{Marini_yambo_2009, Sangalli_2019} was applied.

Yambo requires a highly accurate calculation of the electronic ground state of the system and is directly interfaced with Quantum Espresso (QE)\cite{Gianozzi_quantum_espresso_2009}. Therefore, single-point calculations for the selected AIMD snapshots were performed using the plane-wave DFT code QE. As for the previous DFT calculations, the PBE exchange-correlation functional was employed, whereas Optimised Norm-Conserving Vanderbilt pseudopotentials were used. A plane-wave cutoff of 60\,Ry with an increased $7 \times 7 \times 1$ k-point mesh using the Monkhorst-Pack scheme was applied. The converged single-point calculations and the corresponding single-particle wave functions were then used as starting point to perform a non-self-consistent field calculation, including an increased number of unoccupied bands. For the water-covered Au(110) surface, which contained altogether 1076 electrons, a total of 750 bands was considered. This number of bands ensures that enough states above the Fermi level are included, which is necessary for the subsequent RAS calculation via the Yambo code.

Finally, Yambo calculations were performed to obtain the RA spectrum in an energy range of 0 to 5\,eV. A real-space cutoff was selected such that only the topmost four layers of the Au slab (including the reconstructed top-layer) contributed to the RA spectrum.
As evident from eqn.~(\ref{eqcomras_A}) and ~(\ref{eqcomras_B}), the computational RA spectra have to be normalised by the dielectric function of the bulk material. For this purpose, an experimental bulk dielectric function was applied \cite{Pogodaeva_dielectric_function_six_metals_2021}.

It has to be pointed out that RAS calculation for the here presented system sizes became only possible by using a current Yambo code version with enhanced parallelisation. We therefore ported an optimised RAS module to the public development branch of the latest Yambo version (Yambo 5.1)~\cite{yambo_surf_github}. This module currently features the random phase approximation without local field effects (IP-RPA) to calculate dielectric functions. In the IP-RPA method, the imaginary part of the dielectric function of a slab can be computed as 

\begin{equation}
\label{eqalpha}
\begin{aligned}
\mathrm{Im}[\varepsilon_{s}^{xx}]&=\frac{8\pi^2 e^2}{m^2\omega^2AL}\sum\limits_{\mathrm{\mathbf{k}}}\sum\limits_{v,c}|P_{v\mathrm{\mathbf k},c\mathrm{\mathbf k}}^{x}|^{2}\\& \times \delta(E_{c\mathrm{\mathbf k}}-E_{v\mathrm{\mathbf k}}-\hslash \omega),
\end{aligned}
\end{equation}

where $P_{v\mathrm{\mathbf k},c\mathrm{\mathbf k}}^{x}$ are transition matrix elements, $\omega$ the angular frequency of the light, while $A$ and $L$ correspond to area and height of the cell. These matrix elements consist of the momentum operator $\mathrm{\bf{p}}$ and the commutator containing the non-local elements of the pseudo-potentials $V^{nl}$, \textit{i.e.}, $\mathrm{\bf{p}}+i[V^{nl},\,\mathrm{\bf{r}}]$. 
To calculate the RA spectrum stemming from a specific surface structure on a slab, rather than including the whole slab, a so-called real-space cutoff can be introduced~\cite{Hogan_2003}. By employing such a cutoff $\theta$, the transition matrix elements transform into

\begin{equation}
\Tilde{P}_{v\mathrm{\mathbf k},c\mathrm{\mathbf k}}^{x}=
\big<v\mathrm{\mathbf k}\big|\theta(z)[\mathrm{\bf{p}}^{x}+i[V^{nl},\mathrm{\bf{r}}^{x}]]\big|c\mathrm{\mathbf k}\big>,
\label{eqcutoffP}
\end{equation}
and we can rewrite Eq.~(\ref{eqalpha}) as
\begin{equation}
\begin{aligned}
\mathrm{Im}[\varepsilon_{s}^{xx}]&=\frac{8\pi^2 e^2}{m^2\omega^2AL}\sum\limits_{\mathrm{\mathbf{k}}}\sum\limits_{v,c}[P_{v\mathrm{\mathbf k},c\mathrm{\mathbf k}}^{x}]^{\ast}\Tilde{P}_{v\mathrm{\mathbf k},c\mathrm{\mathbf k}}^{x} \\
&\times \delta(E_{c\mathrm{\mathbf k}}-E_{v\mathrm{\mathbf k}}- \hslash \omega). 
\label{eqcutoffalpha}
\end{aligned}
\end{equation}

This updated implementation, allowing for a significant speedup and larger system sizes, will become part of one of the next Yambo releases.

\section{Results and discussion}

\begin{figure*}[h!]
    \centering
    \includegraphics[width=\textwidth]{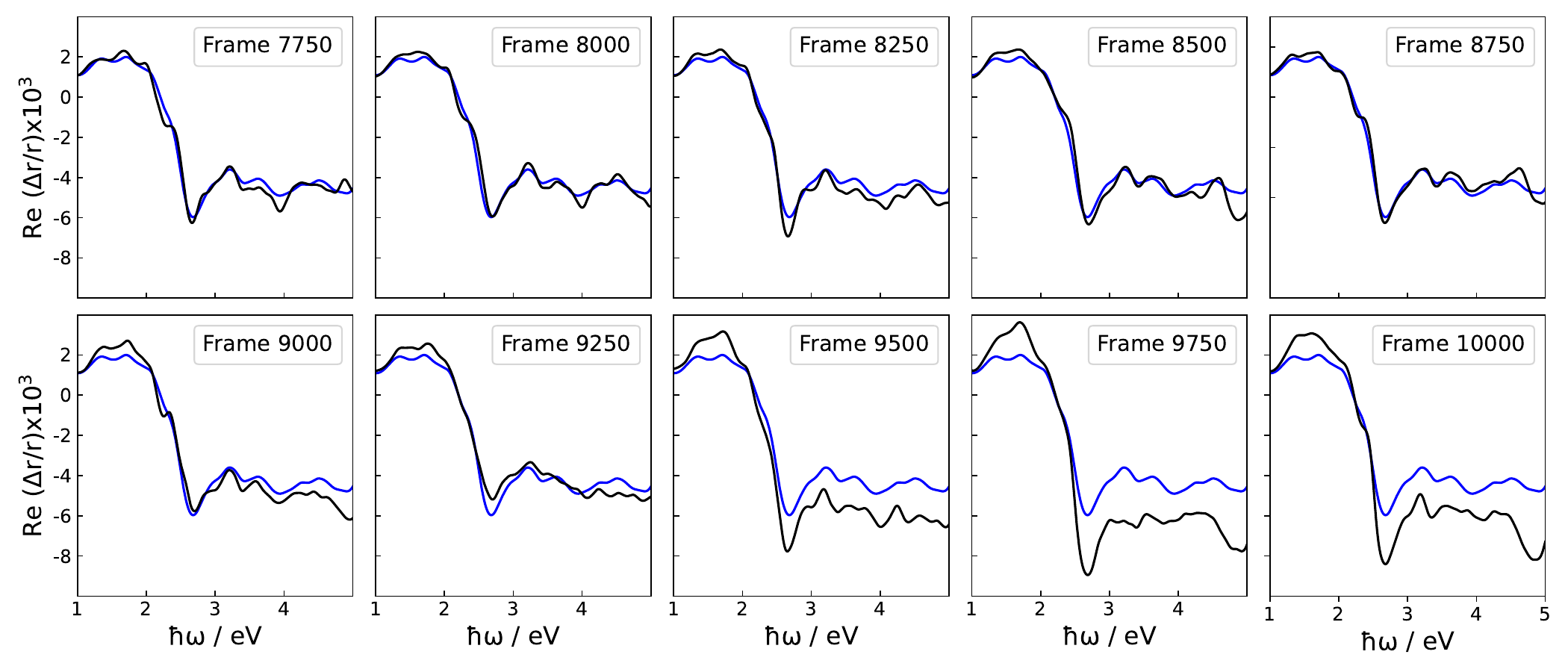} % MM: renamed to a more instructive filename
    \caption{RA spectra of the Au(110) slab in contact with water obtained for configurations extracted from different time frames of the AIMD simulation (black) as compared to the resulting average spectrum (blue).}
    \label{fig:Au110ras-frames}
\end{figure*}

For a typical RAS experiment with a commercial spectrometer, the measured spectrum corresponds to the anisotropy signal averaged over a macroscopic surface area (several mm squared) and integrated over a time of at least 10\,ms. To investigate how the dynamics of the solid-liquid interface of a nm-sized supercell impact the spectroscopic response, we first computed RA spectra along a 2.5\,ps-long part of an MD trajectory of the Au surface in contact with water. The spectral response corresponding to the different MD snapshots and the resulting average are depicted in Figure~\ref{fig:Au110ras-frames}. All spectra show elements of the experimentally observed main features of the reconstructed Au(110) surface, in particular the pronounced negative intensity at 2.7\,eV \cite{Mazine_electrochemical_RAS_Au110_1999, Smith_electrochemical_oxidation_Au110_RAS_2007, Guidat_EC-RAS_review_2023}. Interestingly, the single spectra show certain differences to the time-average, especially in the high-energy part beyond 2.5\,eV (see Fig.~\ref{fig:Au110ras-frames}). It can also be observed that larger deviations are conserved over a longer time period, here for frames 9,500 to 10,000, i.e. over a time-period of at least 250\,fs.

\begin{figure*}[h!]
    \centering
    \includegraphics[width=0.98\textwidth]{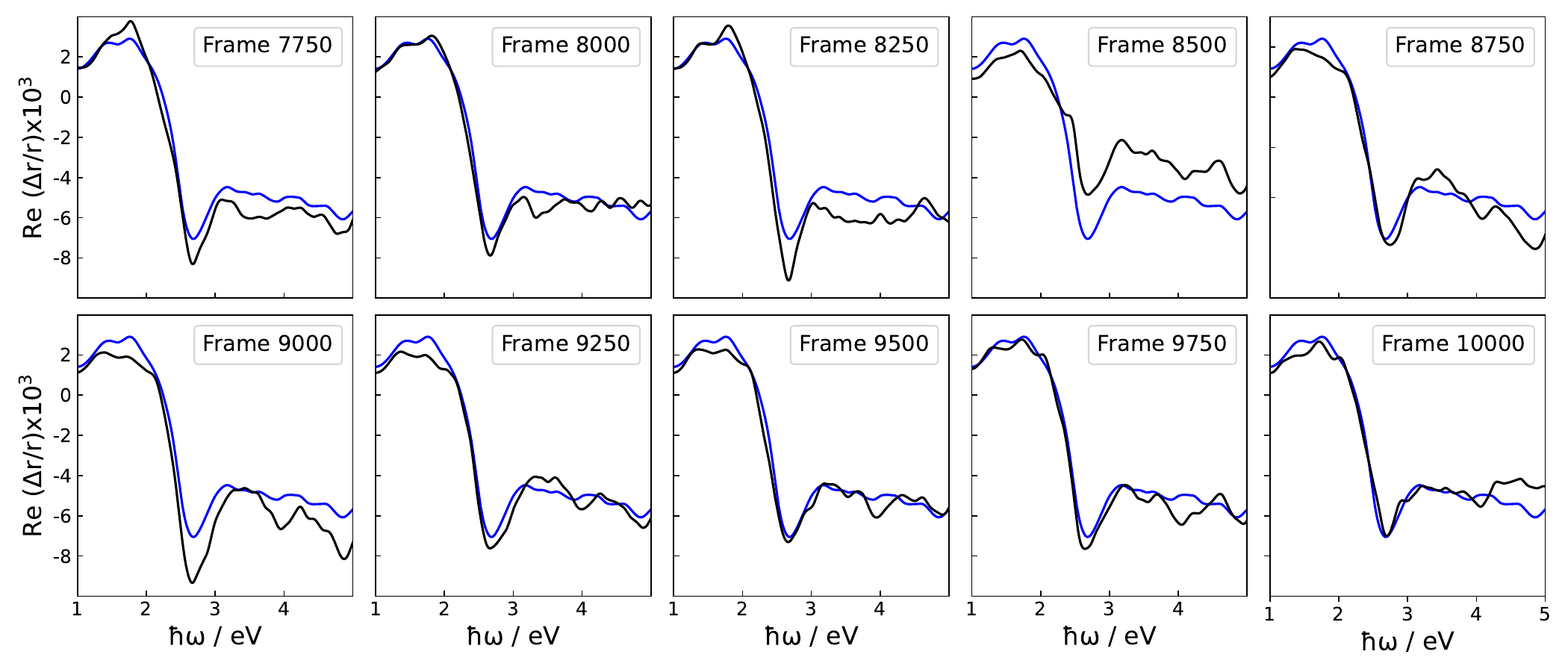}
    \caption{RA spectra obtained for configurations extracted from different time frames of the water-free AIMD simulation (black) of the Au(110) slab as compared to the resulting average spectrum (blue).}
    \label{fig:Au110-nowater-frames}
\end{figure*}

\begin{figure}[h!]
    \centering
    \includegraphics[width=.5\textwidth]{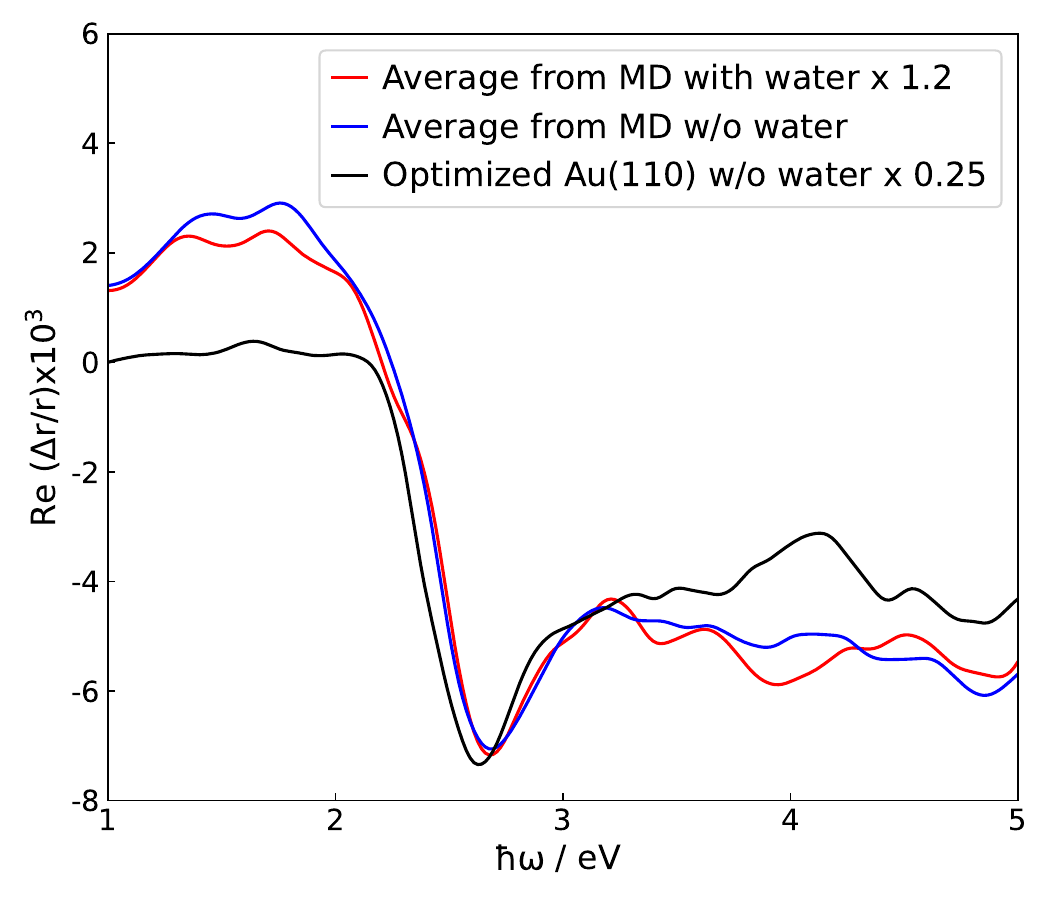} 
    \caption{RA spectra of Au(110) missing row-reconstruction slab in vacuum (black) and averaged spectra for plain surface (blue) and the surface in contact with water (red) from the AIMD runs. Note that the RA spectra are rescaled to match the intensity of the main peak of the water-free MD average for better comparability.}
    \label{fig:Au110ras-comp}
\end{figure}

To further elaborate on the fluctuations induced by the water layer, a comparative MD simulation without water was performed and the resulting RA spectra were investigated. Here, the MD trajectory of the plain Au(110) missing-row reconstruction was also analyzed with respect to the time evolution of the RA spectra (see Fig.~\ref{fig:Au110-nowater-frames}).  Also here, a persistence of larger deviations from the average can be observed, in this case from frames 8,250 to 9,000, which translates to a time-span of 375\,fs. The reason for these deviations is most likely the characteristic frequencies of the Au surface represented in the MD trajectory. This emphasises that computational RA spectra from single arbitrary snapshots of an MD trajectory will typically not give the A comparison of MD-averaged spectra with and without water indicates rather small differences, which are most pronounced in the intermediate energy range from about 2.5 to 3.5\,eV (see Fig.~\ref{fig:Au110ras-comp}).

A comparison to the RA spectrum of the relaxed, water-free equilibrium structure of the Au(110) missing row reconstruction in vacuum with both average spectra shows significant differences. Overall, the intensity of the averaged RA spectra amounts to only about $1/4$ of the intensity obtained for the geometry-optimised, minimum energy surface displayed in Fig.~\ref{fig:Au110ras-comp}. Interestingly, the low-energy, positive anisotropy feature between 1.0 and 2.2\,eV is much more pronounced in the case of the average spectra. A comparison of the averaged spectra with and without water shows that the water-decorated surface leads to a spectrum with an intensity further reduced by about 20\%, indicating a slightly reduced ordering of the Au surface. While most spectral features are almost conserved, the spectrum from the slab with water shows more structure in the energy range around 4\,eV.

The observed intensity decrease for the MD-averaged spectra can be understood as a consequence of the surface dynamics, which makes the originally present anisotropy less pronounced, thus resulting in a reduced RA signal. This is an important finding, as differences between experimental and calculated spectra are often interpreted as stemming from not fully-covered areas on the surface. Our findings, however, strongly indicate that such differences may also be a consequence of neglecting the surface dynamics: While the geometry-optimised surface structure at 0\,K represents ideally the absolute minimum energy configuration of the energy hypersurface with respect to the atomic coördinates, the kinetic energy in the MD simulation will drive the system out of this minimum and can therefore reduce anisotropies of the structure.
 
A remaining question is now the origin of the differences between the averaged RA spectra of the Au surface with and without water. These may either be due to a direct contribution of the water molecules to the RA spectra or due to an indirect impact via a water-induced alteration of the Au surface, {\textit{i.e.}} via changes in structure and dynamics of the interface. To resolve this issue, the RA spectra of the water containing structure were recalculated by applying the real space cutoff such that potential contributions of water molecules were excluded.

\begin{figure}[h!]
    \centering
    \includegraphics[width=.5\textwidth]{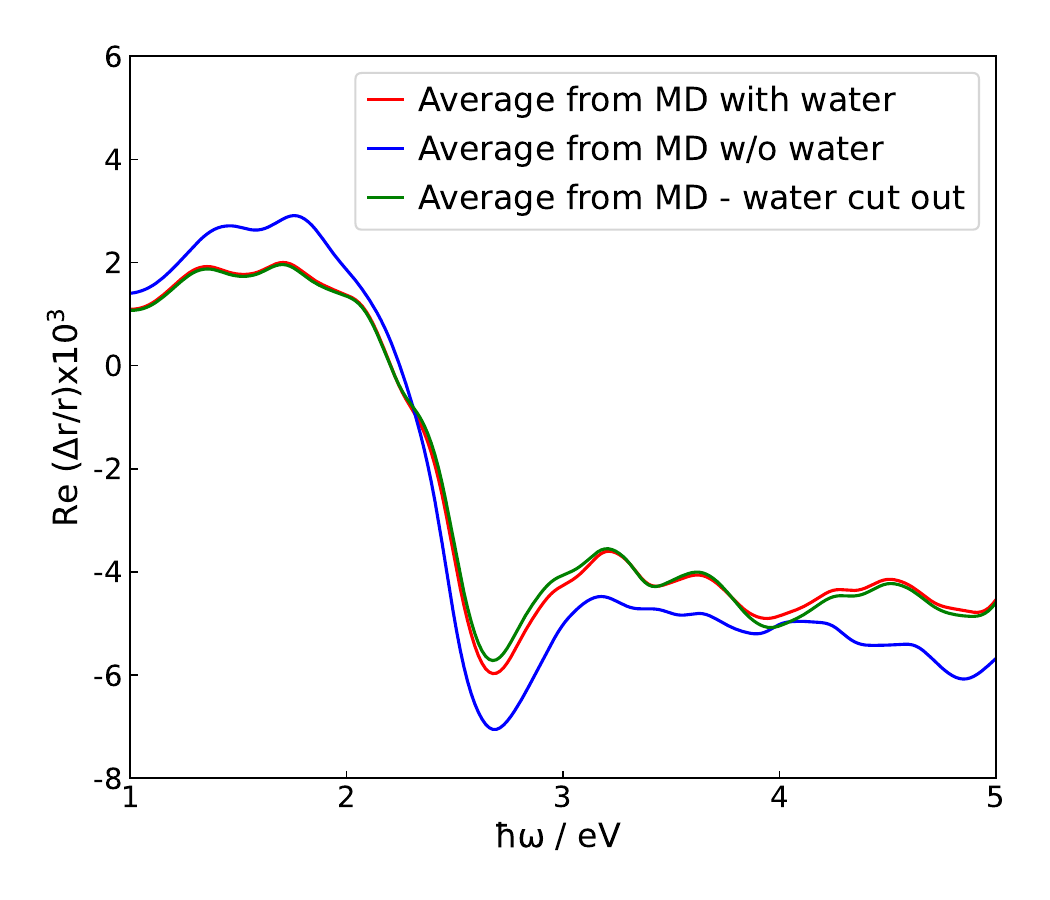}
    \caption{Comparison of averaged spectra for the MD simulation including water. The red curve considers the water layers also for the RAS calculation whereas for the green curve the water contribution is cut out by the real-space cutoff. The blue curve depicts the average spectrum of the MD simulation without water.}
    \label{fig:Au110ras-comp2}
\end{figure}

As can be inferred from Fig.~\ref{fig:Au110ras-comp2}, the average RA spectra obtained for the MD simulation including water is almost independent of the direct water contribution, similar to what was found for a static calculation of a semiconducting system in contact with water \cite{May_water_adsorption_calc_RAS_2018}. This does, on the other hand, mean that the differences between the average spectra obtained for the MD simulation with and without water are an evidence of the impact of the water molecules on real-space and electronic structure of the gold surface. The latter one can also originate from the electric field inducing a linear electro-optic effect \cite{May_water_adsorption_calc_RAS_2018}. In other words, the difference between the average spectra of the two MD simulations appear to be a signature of the solid-liquid interface.

Finally, it has to be pointed out that the here presented model system does not perfectly agree with experimentally observed data of the full electrochemical system \cite{Mazine_electrochemical_RAS_Au110_1999, Smith_electrochemical_oxidation_Au110_RAS_2007, Smith_Au110_electrochemical_RAS_2016, Guidat_EC-RAS_review_2023}. This, however, is likely to be a consequence of the limited system size as well as the absence of electrolyte ions. For a true comparison to experiment, an increased number of Au layers might have to be investigated. Yet, the focus of the present work lies in emphasising the impact of the surface dynamics on the RA signal, which is already evident for the here presented system size.

\section{Conclusion and Outlook}

In this paper, we investigated the impact of structural variability in the time-domain on reflection anisotropy spectroscopy by computing RA spectra along an MD trajectory of the Au(110) surface, featuring the frequently observed missing-row reconstruction with and without considering the presence of water. In both cases the dynamically changing structure of the Au surface is found to result in a significant reduction of the observed optical anisotropy, while at the same time the main RAS peak becomes more pronounced. This finding indicates that surface and interface dynamics are important when RA spectra are analysed with respect to computational data. Indeed, differences in overall signal intensity between experimental and computed RA spectra are frequently attributed to macroscopic properties of the surface, whereas a microscopic origin in terms of surface/interface variability was so far not considered. Furthermore, the observed differences in the RA spectra between the plain surface and the water-containing setup show that the presence of water molecules changes the surface/interface structure in a well-ordered manner. This points to the high sensitivity of RAS with respect to the solid-liquid or solid-electrolyte interface (SEI), while at the same time it becomes evident that for the interpretation of such interfaces, static calculations that only consider the geometry-optimised, minimum energy structure will not necessarily give the full picture. Finally, we want to emphasise the potential of RAS as a method to study long-standing problems in applied electrochemistry, such as the SEI growth in batteries. When single-crystalline model systems can be prepared, RAS is capable of deciphering the processes that are determininig the initial state of the SEI growth, which are still not well understood in many battery systems. With the more complex electrolytes including large molecules, however, this will be a challenging undertaking. Yet, also for aqueous electrochemistry, a closer approximation of real electrochemical systems will require the analysis of MD trajectories that include both ions and applied potentials \cite{Zhang_electrochemical_systems_finite_field_MD_2020}.

\begin{acknowledgments}
This work was funded by the German Research Foundation (DFG) under project number 434023472 and the Cluster of Excellence EXC 2154 / ``Post-Li Storage'', project number 390874152. The German Bundesministerium für Bildung and Forschung (BMBF), supported this work with the project ``NETPEC'' (No.~01LS2103A). The authors acknowledge support by the state of Baden-Württemberg through bwHPC and the German Research Foundation (DFG) through grant no INST 40/575-1 FUGG (JUSTUS 2 cluster). Part of the simulations were performed on the national supercomputer Hawk at the High Performance Computing Center Stuttgart (HLRS) under the grant number SPECSY/44227. The authors thank Conor Hogan and Davide Sangalli for assistance with re-implementing the RAS module in Yambo.
\end{acknowledgments}

\section*{Data availability statement}

The AIMD trajectories supporting this study will be made openly available on the NOMAD repository at https://doi.org/.... (To be published alongside the manuscript.)

\end{document}